\documentclass[aps,prl,twocolumn]{revtex4}
\newcommand{\be}{\begin{equation}}
\newcommand{\ee}{\end{equation}}

\begin{document}


\title{Flavor Democracy: On Which Basis?}


\author{Tongu\c{c} Rador}
\email[e-mail:]{rador@gursey.gov.tr}
\affiliation{Feza G\"{u}rsey Institute,\\ Emek Mah. No:68, \c{C}engelk\"{o}y, 81220, \.Istanbul, Turkey}

\date{\today}

\begin{abstract}
I argue that flavor democracy should not be considered as a "mere" assumption, but rather a
fact that is inherent where there is  hierarchy of quark masses. Briefly the crux of the
argument is the ambiguity of defining a basis when one introduces a mass matrix. That is
there is a degree of freedom of also defining a basis other than the weak eigenbasis with
respect to which one can write down mass matrices. Since the ultimate aim is to diagonalize
to the mass eigenbasis this is physically equivalent. But not necessarily so for the human eye.
\end{abstract}
\maketitle

\section{Introduction}

The physics of flavor is difficult. There are many open questions and the general hope is
that the upcoming experiments might shed some light onto this now-obscure part of particle
physics. The major issue is to undesrstand the Cabibbo-Kobayashi-Maskawa mixing matrix (CKM) of
quarks in the charged weak current as well as the hierarchy in the quark masses. While
the two are obviously related the efforts to find quark mass matrices
and the resulting CKM  is beset with difficulties mainly because the problem is infact
infested with many-to-one relations. One reason (albeit rather overlooked) for this is the
fact that one can define another physically equivalent mass matrix, by unitary transformations
of the quark fields, that might look like anything but the original one. The possibilities are
not only these: one could also write down mass matrices in a basis where there is an ad-hoc
CKM (not obviously the final physical one). These considerations clearly
do not contain new physics but aim at somewhat simplifying a problem that is already rather
difficult. Sticking to the weak eigenbasis to write down mass matrices might actually be
helping the problem remain difficult. The idea here is similar to Legendre of Fourrier transformations
of differential equations: although the physics remain the same the problem "look" simpler
when transformed. In this short letter I will try to demonstrate that there are certain priviledged
basises on which the problem of quark mass matrices and the CKM look simpler.

\section{The Argument}

Consider the "physical" charged electroweak and mass lagrangian:
\be
W_{\mu}\bar{U}\gamma^{\mu}(1-\gamma^{5})\;K\;D+\bar{U}M_{U}^{diag}U+\bar{D}M_{D}^{diag}D\;\;.
\ee
\noindent Here $K$ is the physical CKM and the diagonal quark mass matrices contain the physical quark
masses.
Now let us remember that there is a $SU(3)$ matrix such that its elements have equal norm
\be\label{eq2}
K_{d}\equiv\frac{1}{\sqrt{3}}\left[
\begin{array}{ccc}
e^{i\pi/6} & -e^{-i\pi/6} & 1\\
-e^{-i\pi/6} & e^{i\pi/6} & 1\\
1 & 1 & i\\
\end{array}\right]
\ee

This matrix is defined modulo permutations of its rows(columns) and it can be sandwiched between
two different diagonal $SU(3)$ matrices which would introduce 4 arbitrary phases.

Now consider rotating the quark fields  by matrices similar to the one above.
Unless there are magical cancellations the largest (up-down) quark mass will infest the resulting
(up-down) mass matrices. Also for the CKM the hierarchy will disappear.
This transformation consequently
will switch to a basis in which the mass matrices are somewhat (as far as the norms are concerned)
democratic and an ad-hoc CKM that with hierarchy smeared. It
then follows that "flavor" democracy is not a "mere" assumption: it does exist to some extend
in some basis given there is a large hierarchy of quark masses. The physical meaning of this
internediate basis may be non-existent. Since the ultimate aim is to go to the mass eigenbasis,
it is to be considered as a do-nothing-but-for-something.

This point of view has the following advantage, since in the democratic basis only the largest
quark masses will dominate the mass matrices the only parameters that would remain if not to
be considered a "perturbation"  are the phases of the quark mass matrices. There are four phases
in a norm-democratic mass matrix in one (up-down) sector. One is coming from the Nuyts' theorem (J. Nuyts, Phys. Rev. Lett.{\bf{26}}, 1604, (1971).), that is the
non-hermitean part is a multiple of identity (this is true for the full mass matrix with up and down
quarks combined) and the other three are coming from the hermitean
part of which two (separately for up and down sectors) may be transformed away by the
definition ambiguity (four phases) of the ad-hoc CKM. Thus
the final form of a mass matrix in the democratic basis (ignorig norm perturbations) is
\be\label{eq3}
M^{democ}=m_{3}\left[
\begin{array}{ccc}
e^{i\phi} & e^{i\chi} & 1\\
e^{-i\chi} & e^{i\phi} & 1\\
1 & 1 & e^{i\phi}\\
\end{array}\right]
\ee
\noindent The location of $\chi$ is arbitrary and introduces a discrete definition ambiguity.
Also although we know that experimentally $ArgDet(M)<10^{-10}$ we keep the Nuyts' phase here nevertheless
for completeness.

\section{The choices for the ad-hoc CKM}
The crucial matter is then to pick an ad-hoc CKM in the democratic basis. It is clear that the
matrix in Eq.\ref{eq2} is not ultimately a choice because we argued about rotating the original CKM with it
in the first place to go to the democratic basis. But our aim can seldom be to recover the
"original" CKM via this route. This is impossible. The idea is to find a CKM via this method
that would serve as a solution around which mass ratio effects will be considered a
perturbation. What is interesting is that this solution might have physical relevance
for flavor physics.

Within the idea of democracy the choices for the ad-hoc CKM are not limited to the $K_d$ defined
in Eq.\ref{eq2}, since unitraity tends to frustrate democracy. But we can still do as much as we can.
For example let us consider the following ad-hoc CKM

\be{\label{eq4}}
\frac{1}{3}\left[
\begin{array}{ccc}
-1 & 2 & 2\\
2 & -1 & 2\\
2 & 2 & -1\\
\end{array}\right]
\ee

This matrix is absolutely out of hierarchy thus I consider it to be democratic. Accidentally
this matrix commutes with the democratic mass matrix of \ref{eq3} if $\chi$ and $\phi$ are taken
to be zero. Thus the "would-be physical" CKM of this toy model is the unit matrix. This example demonstrates
that the phases $\chi$ can be used to explore the less (wrt to a unit matrix) hierarchical results.

One can actually generalize the idea of democratic unitary matrices by defining matrices of
the form,

\be
\left[
\begin{array}{ccc}
a & b & c\\
b & a & d\\
c & d & a\\
\end{array}\right]
\ee

The case with $a=b=c=d$ ans with $b=c=d$ are given in Eq.\ref{eq2} and Eq.\ref{eq4} respectively.
For other solutions, when one also would like to introduce non-trivial phases,the elements should not
differ by large order of magnitudes since one indeed solves a system of equation without
small or large numbers. This idea can be further generalized but I do not wish to get in
detail here since this is not a full research article but rather a rapid communication.

\section{A toy Application}

Let us consider a toy application. Take the ad-hoc CKM to be $K_{d}^{1/3}$ (this is within
the idea of democratic unitary matrices) and the mass matrices to be just composed of 1's.
There is one caveat here. The eigenvalues of the mass matrix are $\{ 0,0,3\}$ and one has
to face the null space redundancy for choosing the diagonalization matrices. However we
can still invoke democratic point of view and say that the zero mass states share the
null-space in a "democratic" way that is they both take null space vectors rotated with $\pi/4$.
The resulting CKM has the following form

\be
\left[
\begin{array}{ccc}
0.98 & 0.20 & 0.03\\
0.20 & 0.98 & 0.10\\
0.03 & 0.10 & 0.99\\
\end{array}\right]\;\; \rm{and}\;\; \delta\approx \pi/2
\ee

There was fine tuning of no parameters, yet the Cabibbo angle is well near the physical
value and there is some hierarchy in the third columns.

\section{Conclusion}

The argument here is that there is always a basis in which the mass
matrices will be near norm-democratic. The point however is that the basis with repect to
which these mass matrices are realized may not be the weak eigenbasis. The democratic idea somewhat
reduces the possible choices for mass matrices and the degrees of freedom seem
to have shifted to the choice of the ad-hoc CKM. However the idea of democracy is still present
there because in the democratic basis the "physical" CKM will have its hierarchy smeared. I
believe the democratic basis defined in this way, makes the problem easier in an interesting
way. Since within the idea of there seems to be a finite amount of reasonnable
choices for the ad-hoc CKM.

I must apologize for having omitted references due to the rapid communication nature of this
letter.
\end{document}